\documentclass{elsart}
\usepackage[square,comma]{natbib}
\usepackage{graphicx}
\usepackage{pxfonts}
\usepackage{amssymb}

\journal{}

\begin{document}

\begin{frontmatter}

% Title, authors and addresses

% use the thanksref command within \title, \author or \address for footnotes;
% use the corauthref command within \author for corresponding author footnotes;
% use the ead command for the email address,
% and the form \ead[url] for the home page:
% \title{Title\thanksref{label1}}
% \thanks[label1]{}
% \author{Name\corauthref{cor1}\thanksref{label2}}
% \ead{email address}
% \ead[url]{home page}
% \thanks[label2]{}
% \corauth[cor1]{}
% \address{Address\thanksref{label3}}
% \thanks[label3]{}

\title{A simple method for timing an XFEL source to high-power lasers}

% use optional labels to link authors explicitly to addresses:
% \author[label1,label2]{}
% \address[label1]{}
% \address[label2]{}

\author{Gianluca Geloni,}
\author{Evgeni Saldin,}
\author{Evgeni Schneidmiller}
\author{and Mikhail Yurkov}

\address{Deutsches Elektronen-Synchrotron (DESY), Hamburg,
Germany}

\begin{abstract}

We propose a technique, to be used for time-resolved pump-probe
experiments, for timing an x-ray free electron laser (XFEL) to a
high-power conventional laser with femtosecond accuracy. Our
method takes advantage of the same electron bunch to produce both
an XFEL pulse and an ultrashort optical pulse with the help of an
optical radiator downstream of the x-ray undulator. Since both
pulses are produced by the same electron bunch, they are perfectly
synchronized. Application of cross-correlation techniques will
allow to determine relative jitter between the optical pulse (and,
thus, the XFEL pulse) and a pulse from an external pump-laser with
femtosecond resolution. Technical realization of the proposed
timing scheme uses an optical replica synthesizer (ORS) setup to
be installed after the final bunch-compression stage of the XFEL.
The electron bunch is modulated in the ORS setup by an external
optical laser. Subsequently, it travels through the main
undulator, and produces the XFEL pulse. Finally, a powerful
optical pulse of coherent edge radiation is generated as the bunch
passes through a long straight section and a separation magnet
downstream of the main undulator. Our study shows that at a
moderate (about $10 \%$) density modulation of the electron bunch
at the location of the optical radiator allows production of high
power x-ray and optical pulses. Relative synchronization of these
pulses is preserved by using the same mechanical support for both
x-ray and optical elements transporting radiation down to the
experimental area, where single-shot cross-correlation between
optical pulse and pump-laser pulse is performed. We illustrate the
potential of the proposed timing technique with numerical examples
referring to the European XFEL facility.

\end{abstract}


\end{frontmatter}

% main text

\clearpage

\section{\label{sec:intro} Introduction}

Time-resolved experiments are used to monitor time-dependent
phenomena. In typical pump-probe experiments a short probe pulse
follows a short pump pulse at some specific delay. Femtosecond
capabilities have been available for some years at visible
wavelengths \cite{Zewail}. However, there is a strong interest in
extending these techniques to x-ray wavelengths, where one could
directly probe structural changes with atomic resolution. This
goal will be achieved with the realization of x-ray free electron
lasers (XFELs) \cite{XFEL,SLAC,SCSS}. In their initial
configurations, XFELs will produce radiation pulses with duration
of about a hundred femtosecond, which will allow time-resolved
studies of transient structures of matter on the time-scale of
chemical reactions.

One of the main technical problems for the realization of
pump-probe experiments is the relative synchronization of
radiation pulses from XFEL and optical laser. Both x-ray and
optical pulses are subject to time-jitter. In an XFEL, radiation
is produced by an electron bunch travelling through an undulator.
The budget for time jitter of the electron bunch starts to
accumulate from the photo-injector laser system. Extra source of
jitter is constituted by fluctuations of the electron energy,
which transform to time jitter in magnetic bunch-compressors. The
pump-laser itself has an intrinsic time jitter caused e.g. by
mechanical vibrations in the laser oscillator or electrical-noise
in the stabilization electronics.

There exists a tendency for straightforward solution of the
problem by means of implementation of a precise synchronization
system at XFELs, aiming all clocks and triggers within the
facility to be perfectly synchronized \cite{GS1,GS2,GS3}. Once
this is done, temporal resolution of pump-probe experiment will be
defined by the synchronization system plus intrinsic time-jitter
of the pump-laser system. However, precise synchronization on a
kilometer-scale facility is a rather challenging task, and it is
not clear, at the moment, what will be the overall practical
accuracy limit for the synchronization of XFEL pulses to pulses
from external lasers.

Approaches based on generation of two coherent radiation pulses of
different colors by the same electron bunch passing through two
distinct insertion devices have been proposed in \cite{TC1,TC2}.
Intrinsic feature of these schemes is an ideal mutual
synchronization of the radiation pulses. With careful design of
the optical transport system from source to sample it is possible
to reach femtosecond level of synchronization. One of the schemes,
using far infrared and VUV radiation pulses is being
experimentally realized at FLASH, the free-electron laser in
Hamburg \cite{fir-pp}.

Recently, an alternative concept has been proposed for realization
of pump-probe experiments with XFEL pulses and powerful optical
pulses from external lasers \cite{BILL,fel07-TUPPH013}. The basic
idea is to shift the attention from the problem of absolute
synchronization to the problem of measurement of relative jitter
between x-ray and optical pulses. If a relative delay is known for
each pump-probe event, a sorting of results according to measured
delay times gives the required experimental outcome. In this paper
we propose a concept of time-arrival monitor allowing measurement
of a relative delay between XFEL and optical pulses on a
femtosecond time scale. Our scheme makes use of the same electron
bunch to produce both an XFEL pulse and a powerful laser-like
(bandwidth-limited and diffraction-limited) optical pulse. The
latter is generated downstream of the x-ray undulator, and will be
naturally synchronized to the XFEL pulse. Relative synchronization
is preserved by using the same mechanical support for optical
elements transporting x-ray pulse and optical pulse down to the
experimental area. There, single-shot cross-correlation between
the optical pulse and the pump-laser pulse can be performed
yielding, effectively, the temporal delay between XFEL pulse and
pump-laser pulse.

Implementation of the proposed scheme for pump-probe experiments
into the project of the European XFEL \cite{XFEL} is described in
detail. We show that it naturally fits into the project. A
modulation at optical frequency is imprinted onto the electron
bunch using the optical replica synthesizer (ORS) setup
\cite{ORS1} that will be installed after the final beam
compression stage at the energy of 2 GeV. Subsequently, electrons
travel through the main undulator and radiate the XFEL pulse.
Finally, the presence of a long straight section and a separation
magnet downstream of the main undulator serve as optical radiator,
where the modulated electron bunch produces a powerful pulse of
coherent edge-radiation.

The proposed arrival-time monitor can be realized without the
addition of specific hardware in the accelerator and undulator
tunnels, and requires minimal efforts. The applicability of the
present scheme is not restricted to the European XFEL setup. Other
projects, e.g. the LCLS or SCSS facility \cite{SLAC,SCSS}, may
benefit from our proposal too.

\section{Timing system description}

\begin{figure}
\begin{center}
\includegraphics*[width=140mm]{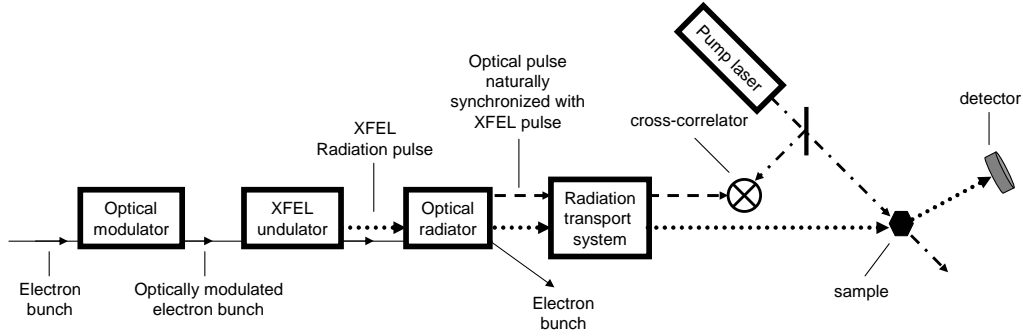}% Here is how to import EPS art
\caption{\label{setup} Scheme for pump-probe experiments with XFEL
pulses and pulses produced by an external laser. Operation of the
scheme is based on production by the electron bunch of an optical
pulse naturally synchronized with the XFEL pulse. Relative
synchronization is preserved by using the same radiation transport
system to the experimental area. The optical pulse is then used to
monitor the delay time between pump and probe by correlating it
with the pump laser.}
\end{center}
\end{figure}

\begin{figure}
\begin{center}
\includegraphics*[width=140mm]{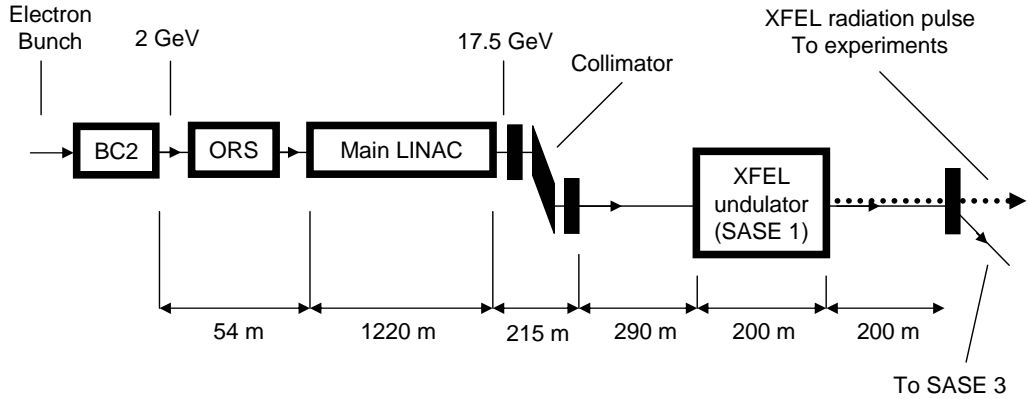}% Here is how to import EPS art
\caption{\label{realscheme} Integration of the time-arrival
monitor in the European XFEL setup makes use of the Optical
Replica Synthesizer (ORS). The ORS is used to imprint density
modulation onto electron beam. Optical radiator is the combination
of straight-section and separation magnet after the SASE 1
undulator which produsese powerful optical
pulse of coherent edge radiation.} \end{center} \end{figure}

A basic scheme of the timing system is shown in Fig. \ref{setup}.
Discussions in this paper focus on design and parameters (see Fig.
\ref{realscheme}) of the SASE 1 line of the European XFEL,
operating in the wavelength range around $0.1$ nm. Main elements
of the timing system are energy modulator (constituted by seed
laser and undulator modulator), dispersion section and radiator.

The energy modulator (see Fig. \ref{ORS}) is located after the
second bunch compressor (BC2), where the electron energy is about
$2$ GeV. A laser pulse at wavelength $\lambda = 400$ nm is used to
modulate the electron energy at the same wavelength. The duration
of the laser pulse is chosen to be about $1$ ps, much longer than
the time-jitter of the electron pulse (a fraction of a picosecond)
to avoid synchronization problems. The energy of the laser pulse
is about $0.5$ mJ. Laser beam is focused onto the electron beam in
a short modulator undulator, with a number of periods $N_w = 5$,
resonant at the seed laser wavelength $\lambda = 400$ nm. Energy
modulation with an amplitude of about $0.5$ MeV is achieved due to
interaction with the electron bunch in the modulator undulator.
Subsequently, the electron bunch passes through the dispersion
section with momentum compaction $R_{56} \simeq 15 \mu$m, where
the energy modulation induces a density modulation at the
seed-laser wavelength. With parameters discussed above, the
density modulation reaches an amplitude of about $5\%$. Following
the dispersion section the bunch is accelerated up to the energy
of $17.5$ GeV in the main accelerator, it passes trough the
collimator system and is directed to the SASE 1 x-ray undulator.

A high-current (5 kA) electron beam is transported through the
XFEL linac, and it is therefore mandatory to include
self-interaction effects in our analysis. During the passage of
the bunch through the accelerator, the initial density modulation
produces an energy modulation due to longitudinal impedance caused
by space-charge fields. If the collimation system is properly
tuned, such energy modulation can induce further modulation in
density. Calculations presented in Section \ref{sec:modu} show
that exploitation of self-interaction mechanisms allow one to
deliver, at the entrance of the SASE 1 undulator, an electron
bunch with a $10 \%$ density modulation, and negligible energy
modulation.

The SASE FEL process in the baseline undulator is not perturbed by
such level of density modulation: fluctuations of the electron
current density serve as input signal for the radiation
amplification process, which develops nearly in the same way as
with an unmodulated electron bunch. As a result, at the exit of
the SASE 1 undulator the electron bunch produces the nominal x-ray
pulse.

Finally, after the SASE 1 undulator, the modulated electron bunch
passes through a long straight section followed by a separation
magnet, which separates the electron bunch from the SASE pulse.
Due to the presence of the combination of straight section and
separation magnet, the electron bunch emits an optical pulse of
coherent edge radiation at the wavelength of the bunch density
modulation. Such an optical pulse carries about $10^{12}$ photons
(order of a microjoule) and is delivered in a bandwidth-limited
and diffraction-limited pulse.

Summing up, combination of an optical modulator after magnetic
bunch compressor BC2 and optical radiator after the baseline
undulator will allow to produce a powerful laser-like optical
pulse at the entrance of the photon beamline. This laser-like
pulse is naturally synchronized with the electron bunch and
thereby with the x-ray pulse. Thus, the problem of measuring the
time-shift between a pump-laser pulse and a (x-ray) probe pulse is
reduced to the problem of measuring the time delay between two
ultrashort optical pulses, which may be solved with standard
techniques.

Contrarily to the x-ray pulse, which is peaked on-axis, the
edge-radiation pulse is peaked in the forward direction at an
angle of a few tens of microradians. The distance from the
separation magnet to the first (x-ray) mirror station is about
$300$ m. According to calculations presented in Section
\ref{sec:rad}, the diameter of the spatial distribution of edge
radiation at this distance will be about $3$ cm, which fits with
the aperture of the photon beamline.

Relative synchronization of x-ray and optical pulses must be
preserved during propagation to the experimental area. The best
way to do so is to use the same mechanical support for both
optical elements transporting the x-ray pulse and optical elements
transporting the optical pulse. For example, optical elements for
the transportation of edge-radiation may be directly assembled on
the x-ray mirrors. After passing through the transport system, the
edge-radiation pulse will reach the experimental area, where a
single-shot cross-correlation measurement with the pump-laser
pulse can be performed. In reference \cite{CROS} the possibility
of extracting the temporal shift between pulses from two
ultrashort laser pulses was experimentally demonstrated. The
method was based on sum-frequency generation in a non-linear
crystal. The intrinsic accuracy of this technique was shown to be
within the femtosecond range. Successful measurements of the
time-offset signal were performed even when one of the two pulses
was very weak (down to $10^{6}$ photons per pulse).

\section{\label{sec:modu} Operation of the optical modulator}

\begin{figure}
\begin{center}
\includegraphics*[width=140mm]{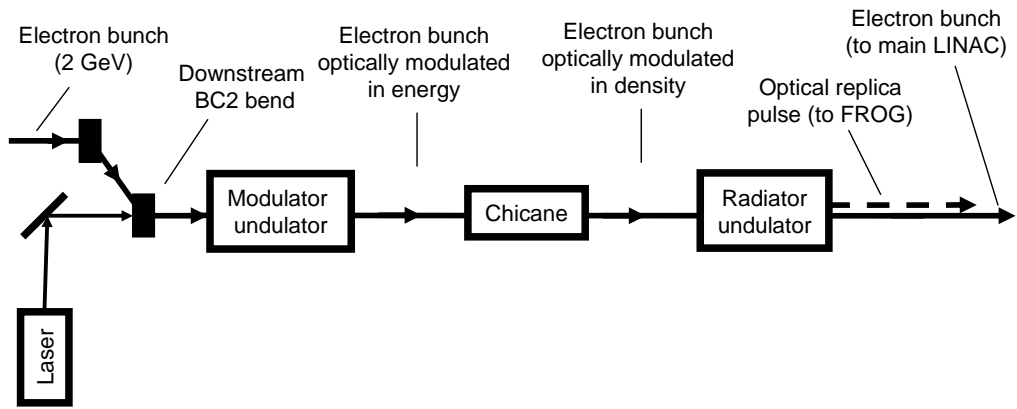}% Here is how to import EPS art
\caption{\label{ORS} The Optical Replica Synthesizer (ORS) will be
used to superimpose density modulation at optical wavelength on
the electron bunch. First, energy modulation is created by letting
a laser interact with the bunch inside a modulation undulator.
Second, a dispersion section (magnetic chicane) induces the
desired density modulation. The radiator undulator, used in
optical-replica ultrashort electron bunch diagnostics, is switched
off during operation of the arrival-time monitor.}
\end{center}
\end{figure}
The modulator to be used in our scheme is the optical replica
modulator \cite{ORS1}, which consists of three elements (see Fig.
\ref{ORS}): an optical seed laser, a modulator undulator and a
dispersion section (magnetic chicane). The radiator undulator
shown in Fig. \ref{ORS}, used in optical-replica ultrashort
electron bunch diagnostics, is switched off during operation of
the arrival-time monitor. The seed laser pulse interacts with the
electron beam in the modulator undulator, which is resonant with
the laser wavelength $\lambda$. As a result, the electron bunch is
modulated in energy. In the following dispersion section the
energy modulation induces a density modulation at the optical
wavelength.

The dispersion section is designed to produce an energy dependence
of the particles path length $\delta z = R_{56} \delta
\gamma/\gamma_0$, where $\delta \gamma$ is the energy deviation of
a particle in units of the rest mass, $\delta z$ is the deviation
from the path length of an electron with nominal energy $\gamma_0$
in units of the rest mass, and $R_{56}$ is the compaction factor
of the dispersion section. $\delta \gamma$ and $\delta z$ may
assume positive or negative values, while $R_{56} > 0$, the
dispersion section being a chicane. Suppose that at the entrance
of the chicane we have an initial energy modulation. Then, in
units of the rest mass, the energy deviation of each particle due
to this energy modulation is given by $(\Delta \gamma)_i
\sin(\psi)$, where  $\psi = \omega[z/v_z(\gamma_0) - t]$ is the
modulation phase, with $v_z$ the longitudinal velocity of a
nominal electron, and $\omega= 2\pi c/\lambda$, $c$ being the
speed of light in vacuum. Here $(\Delta \gamma)_i$ is the
amplitude of our small ($|(\Delta \gamma)_i|/\gamma_0 \ll 1$)
energy modulation taken with its own sign. At the exit of the
chicane we may express the current as $I = I_0[1+ a_i
\cos(\psi)]$. Here $I_0$ is the unmodulated electron beam current,
and $a_i$ is the amplitude of our small ($|a_i| \ll 1$) density
modulation, also taken with its own sign.

Neglecting collective effects, the amplitude of density modulation
at the exit of the chicane, $a_i$, approaches \cite{Czonka}

\begin{eqnarray}
a_i = \frac{R_{56} }{\lambdabar} \frac{(\Delta
\gamma)_i}{\gamma_0} \exp\left[-\frac{1}{2}
\frac{\left\langle(\delta \gamma)^2\right\rangle}{\gamma_0^2}
\frac{R_{56}^2}{\lambdabar^2}\right]~, \label{a1}
\end{eqnarray}
where $\left\langle(\delta \gamma)^2\right\rangle^{1/2}$ is the
rms uncorrelated energy spread of the electron bunch in units of
the rest mass, and $\lambdabar = \lambda/ (2\pi)$ is the reduced
modulation wavelength. In the case of the European XFEL (see
\cite{XFEL}, Fig. \ref{realscheme} and Fig. \ref{ORS}) $\gamma_0
\simeq 4 \cdot 10^3$, corresponding to an energy of $2$ GeV,
$\left\langle(\delta \gamma )^2\right\rangle^{1/2}\simeq 2$,
corresponding to $1$ MeV rms uncorrelated energy spread, $(\Delta
\gamma)_i \simeq 1$, corresponding to $0.5$ MeV modulation, and
$\lambda = 400$ nm, corresponding to the second harmonic of a
Ti:Si laser. A value $R_{56} \simeq 30 \mu$m leads to a modulation
amplitude $a_i \simeq 0.1$. Note that in this case the exponential
suppression factor in Eq. (\ref{a1}) is about $0.98$, and can be
practically neglected.

If collective effects could be neglected during the transport of
the bunch, one could propagate the $10 \%$ initial density
modulation up to the radiator, accounting for velocity bunching
and, possibly, for the presence of a non-zero compaction factor at
the collimation section $R^{(c)}_{56}$. In this case, estimations
would remain within the framework of single-particle dynamics.
Tuning the energy in the optical seed laser pulse, one may easily
achieve a final modulation $a_f=0.1$ at the optical radiator
entrance.

However, collective effects strongly influence the modulation
process in our jitter-monitoring scheme. In other words, the
problem of propagation of the induced beam density modulation
through the setup depicted in Fig. \ref{realscheme} is a problem
involving self-interactions. As the bunch progresses through the
linac, the modulation of the bunch density produces an energy
modulation due to longitudinal impedance caused by space-charge
field. This process is complicated by the fact that, due to the
presence of energy and density modulation, plasma oscillations can
develop. One should account for these facts in order to quantify
energy and density modulation before the collimation section.
Then, in the collimation section, the energy modulation induces
extra-density modulation due to non-zero compaction factor.
Finally, longitudinal space-charge impedance is responsible for
further energy modulation and further plasma oscillations during
the passage of the beam in the main undulator. As a result, the
initial beam modulation, in energy and density, will be modified
by the passage through the setup. In order to study the
feasibility of our scheme one needs to estimate what modifications
will take place.

At the entrance of the accelerator, the density modulation
amplitude is given by $a_i$, and the energy modulation amplitude
by $(\Delta \gamma)_i$. Due to energy modulation, particles
undergo a phase shift, with respect to the phase $\psi$, which is
responsible for a change in the density modulation along the
acceleration section. Also the energy modulation is a varying
function of $z$ because of the presence of longitudinal
space-charge forces. Indicating with $a(z)$ and with $\Delta
\gamma(z)$ the density and energy modulation amplitudes along the
accelerator we may write

\begin{eqnarray}
\frac{d a}{dz} = \frac{1}{\lambdabar} \frac{\Delta
\gamma(z)}{\gamma^3(z)}~,\label{Daiacc}
\end{eqnarray}
where, additionally, the relativistic Lorentz factor of the bunch,
$\gamma(z)$, accounts for the acceleration process. Eq.
(\ref{Daiacc}) can be directly derived from Eq. (\ref{a1})
substituting $a_i$ and $(\Delta \gamma)_i$ with $a(z)$, and
$\Delta \gamma(z)$, neglecting the exponential suppression factor
due to uncorrelated energy spread, remembering that the compaction
factor for a free-space section of length $\delta z$ is given by
$R_{56} = \delta z/(\lambdabar \gamma^2)$, and taking the limit
for $\delta z \longrightarrow 0$.

One can estimate the derivative of $\Delta \gamma(z)$ along the
XFEL linac using results from papers studying microbunching
instabilities like \cite{LSCF}. We can write:

\begin{eqnarray}
\frac{d (\Delta \gamma)}{dz}  = - \frac{4 \pi}{Z_0} a(z)
\frac{I_0}{I_A} \frac{ d |Z|}{d z} ~, \label{modul}
\end{eqnarray}
where $Z(z,\omega)$ is the longitudinal impedance induced by space
charge, $Z_0$ is the free-space impedance, expressed in the same
units of $Z$, and $I_A \simeq 17$ kA the Alfven current.

The longitudinal impedance induced by space charge in free space
was studied in the case of an electron bunch with finite
transverse profile in \cite{ROSE,LWF1}. In that reference we gave
analytical expressions for the impedance in the steady state
limit, which can be easily generalized in the case of adiabatic
acceleration, when $\lambdabar (d \gamma^2/dz) = 2 \lambdabar
\gamma (d \gamma/dz) \ll 1$. The adiabatic acceleration limit can
always be used in our case. Assuming a constant acceleration
gradient $d\gamma/dz \equiv g \simeq 25 \mathrm{m}^{-1}$ (see
\cite{XFEL}), we have $2 \lambdabar \gamma (d \gamma/dz) \simeq
0.01$ for $\gamma = 4\cdot 10^3$, corresponding to the lowest
energy of $2$ GeV, and $2 \lambdabar \gamma (d \gamma/dz) \simeq
0.1$ for $\gamma = 3.5 \cdot 10^4$, corresponding to the highest
energy of $17.5$ GeV, whereas the largest effects due to
longitudinal impedance are expected in the first part of the
acceleration process. Then, using results in \cite{LWF1}, which
are valid for a Gaussian transverse distribution of the electron
bunch, we find that Eq. (\ref{modul}) can be written as

\begin{eqnarray}
\frac{d(\Delta \gamma)}{dz} \simeq  - \frac{a(z)}{\lambdabar
(\gamma_0+g z)^2} \frac{I_0}{I_A} {\exp\left[\frac{\epsilon_n
\beta}{(\gamma_0+g z)^3 \lambdabar^2}\right]
~\Gamma\left[0,\frac{\epsilon_n \beta}{(\gamma_0+g z)^3
\lambdabar^2}\right]}~. \label{modul2}
\end{eqnarray}
Here $\Gamma$ is the incomplete gamma function, $\beta$ is the
average betatron function in the accelerator, and $\epsilon_n$ is
the normalized emittance.

The system of coupled differential equations constituted by Eq.
(\ref{Daiacc}) and Eq. (\ref{modul2}) should be solved with given
initial conditions $a(0)=a_i$ and $\Delta \gamma(0)=(\Delta
\gamma)_i$ at the entrance of the accelerator in order to obtain
density and energy modulation at the entrance of the collimator.
Such system of equations accounts for plasma oscillations of the
electron bunch in the limit for adiabatic acceleration, with the
help of a longitudinal impedance averaged along the transverse
direction. A more detailed analysis of space-charge waves
performed as a function of the transverse coordinates (i.e.
without averaging the longitudinal impedance) is given in
\cite{SCTH}, where we studied the problem of plasma oscillations
for an electron bunch with arbitrary transverse profile going
along a straight section with uniform motion.

We assume an average betatron function of about $\beta = 25$ m
along the main accelerator and a normalized emittance $\epsilon_n
= 1.4$ mm$\cdot$mrad (see \cite{XFEL}). Setting the acceleration
length $d_a \simeq 1220$ m (see Fig. \ref{realscheme}) and $I_0
\simeq 5$ kA, numerical analysis shows that our initial conditions
$a_i \simeq 0.05$ and $(\Delta \gamma)_i \simeq 1$ yield, at the
entrance of the collimator $z=d_c$, $a(d_c) \simeq 0.03$ and
$\Delta \gamma(d_c) \simeq -6$.

The nominal value of the compaction factor of the collimator
$R^{(c)}_{56}$ is set to zero, with possibility of fine tuning
around this value of about $\pm 100 ~\mu$m. Taking advantage of
this possibility and setting $R^{(c)}_{56} \simeq + 50 \mu$m, at
the exit of the collimator one obtains a density modulation $a_c
\simeq -0.1$, whereas the energy modulation remains unvaried
$(\Delta \gamma)_c  \simeq -6$.

\begin{figure}
\begin{center}
\includegraphics*[width=140mm]{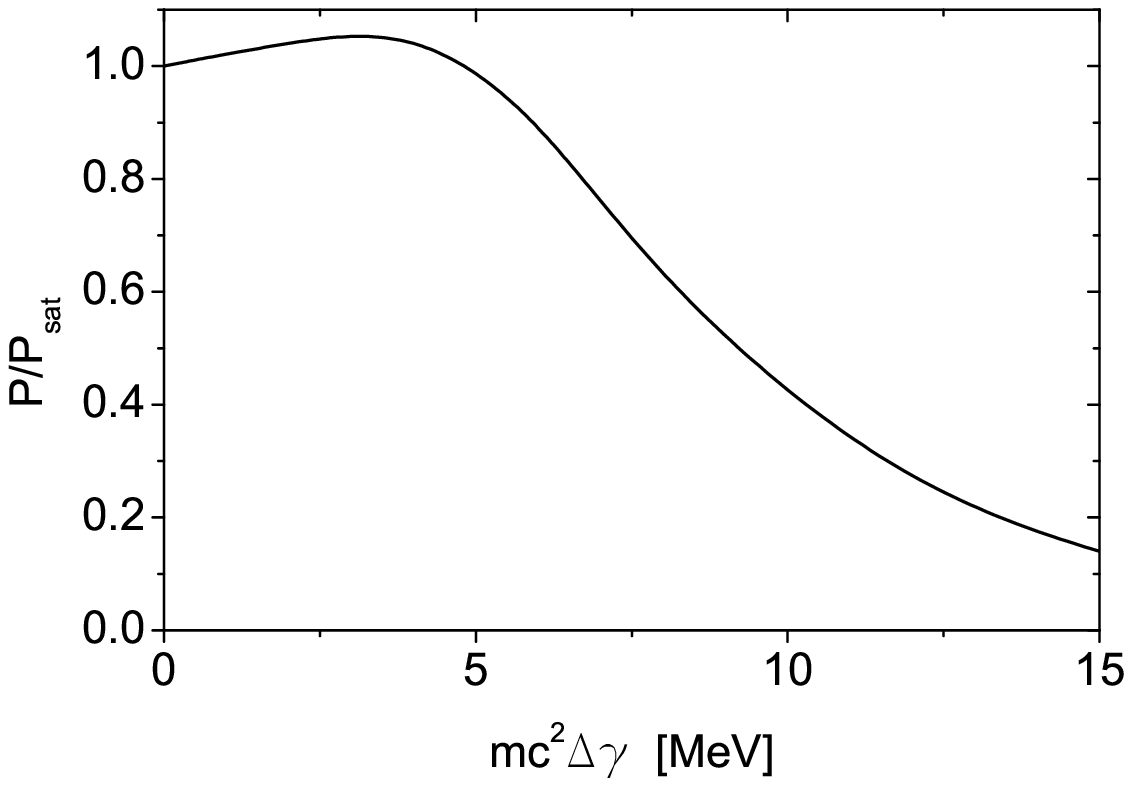}
\caption{\label{fig:psat-enmod} Dependence of the FEL output power
on the energy modulation for undulator SASE 1 at the European XFEL
\cite{XFEL}. Amplitude and period of electron bunch modulation are
10\% and 400 nm, respectively. Radiation power is normalized to
saturation power for an unperturbed bunch. Simulations have been
performed with the code FAST \cite{FAST}.}
\end{center}
%\end{figure} %
%\begin{figure}
\begin{center}
\includegraphics*[width=0.45\textwidth]{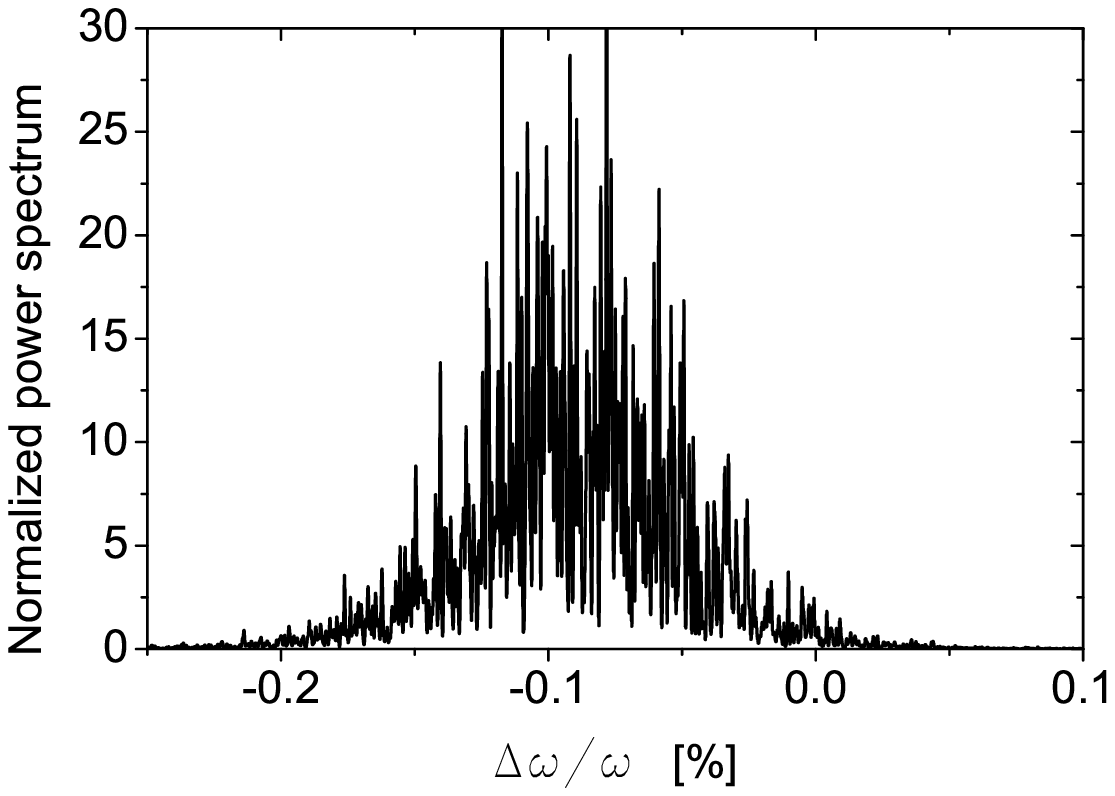}
\includegraphics*[width=0.45\textwidth]{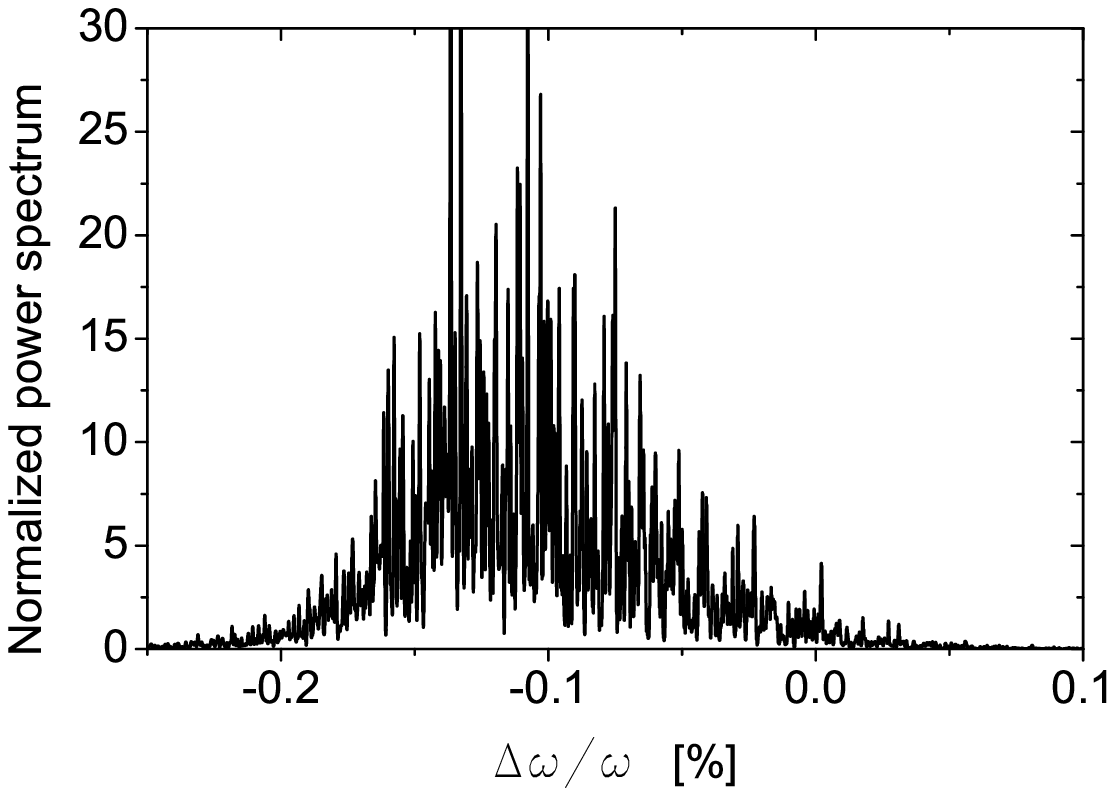}

\caption{\label{fig:spectrum-0-4} Normalized spectrum of the FEL
radiation pulse at zero energy modulation (left plot), and at
$\Delta \gamma mc^2 = 4$~Mev (right plot). Numerical example
corresponds to undulator SASE 1 at the European XFEL \cite{XFEL}.
Amplitude and period of electron bunch modulation are 10\% and 400
nm, respectively. Simulations have been
performed with the code FAST \cite{FAST}.} \end{center} \end{figure}

Further on, energy and density modulation should be propagated
through a straight section followed by the main XFEL undulator.
Propagation can be performed using the same set of equations Eq.
(\ref{Daiacc}) and Eq. (\ref{modul2}), setting $g=0$, using an
energy of $17.5$ GeV and an average value of the betatron function
$\beta = 20 $m. This gives only a correction to the energy
modulation, so that at the entrance of the XFEL undulator, at $z =
d_u$, one still has $a(d_u) \simeq -0.1$, while $\Delta \gamma
(d_u) \simeq -5$. Decrease of the energy modulation is related to
an advantageous initial phase of plasma oscillation at the
entrance of the straight section. Numerical analysis shows that
such an energy modulation, being smaller than the foreseen XFEL
spectral bandwidth $0.08 \%$ \cite{XFEL}, will not alter the XFEL
process (see Figs.~\ref{fig:psat-enmod} and
\ref{fig:spectrum-0-4}).

Similarly as before, the passage in the main XFEL undulator has
the effect of decreasing the energy modulation level, too.
Moreover, although the undulator is shorter than the straight
section preceding it, the effect on the energy modulation is
stronger. In fact, the longitudinal Lorentz factor
$\gamma_z=\gamma/\sqrt{1+K^2/2}$ should be used in the undulator
instead of $\gamma$ (see reference \cite{OURX}). Since $K = 3.3$,
$\gamma^2$ and $\gamma_z^2$ differ of about an order of magnitude,
hence the different influence of the undulator compared with the
straight section. In the undulator, Eq. (\ref{Daiacc}) and Eq.
(\ref{modul2}) are modified to

\begin{eqnarray}
\frac{d a_i}{dz} =  \frac{1}{\lambdabar} \frac{\Delta
\gamma}{\gamma \gamma_z^2}~,\label{DaiaccX}
\end{eqnarray}
and

\begin{eqnarray}
\frac{d\Delta \gamma}{dz} \simeq  - \frac{a_i(z)}{\lambdabar
\gamma_z^2} \frac{I_0}{I_A} {\exp\left[\frac{\epsilon_n
\beta}{\lambdabar^2 \gamma_0 \gamma_z^2}\right]
~\Gamma\left[0,\frac{\epsilon_n \beta}{\lambdabar^2 \gamma_0
\gamma_z^2}\right]}~. \label{modul2X}
\end{eqnarray}
Solving numerically with previously found initial conditions
$a(d_u)$ and $\Delta \gamma (d_u)$, and using $\beta=40 $m, one
finds the energy and density modulation levels at the radiator
entrance, $a_f \simeq -0.1$ and $(\Delta \gamma)_f \simeq -2$.

As a final remark it should be noted that, pending design
finalization, the $R^{(c)}_{56}$ value may be set to any value
from $-1$ mm to $+1$ mm, different from the nominal value
$R^{(c)}_{56}=0$. Depending on this value and, possibly, using the
fine tuning option to increase or decrease the $R^{(c)}_{56}$
value up to $\pm 100 \mu$m, different initial conditions should be
set to obtain acceptable values for $a_f$ and $(\Delta \gamma)_f$.
For example, if $R^{(c)}_{56} = +1$ mm, setting $a_i = 0$ and
$(\Delta \gamma)_i \simeq 0.3$ (corresponding to about $0.15$ MeV)
would yield $a_f \simeq -0.1$ and $(\Delta \gamma)_f \simeq 6$,
corresponding to about $3$ MeV energy modulation, which is
perfectly compatible with our scheme.

It follows that the optical modulator can induce about $10\%$
density modulation at the entrance of the optical radiator, $|a_f|
\simeq 0.1$, and acceptable energy modulation, independently of
the design of the collimation section, without perturbation of the
FEL process in the baseline undulator.

\section{\label{sec:rad} Operation of the optical radiator}

After collimation we deal with an electron bunch modulated in
density at optical wavelength. This wavelength is much larger than
the geometrical emittance of the beam, and in our case the
electron bunch can be considered as a filament with no transverse
dimension nor divergence, as far as optical wavelengths are
concerned. An analysis of the problem in the space-frequency
domain \cite{ARTF} shows that when a filament beam modulated in
density passes through a given trajectory, it produces coherent
radiation very much likely a single electron. In general, one
needs to solve Maxwell's equation for given macroscopic sources. A
paraxial treatment is possible, based on the ultrarelativistic
assumption $1/\gamma^2 \ll 1$. Consider the transverse components
of the Fourier transform of the electric field. They form a vector
$\vec{\bar{E}}(\vec{r},z)$, dependent on transverse and
longitudinal coordinates $\vec{r}$ and $z$. From the paraxial
approximation follows that the envelope $\vec{\widetilde{E}} =
\vec{\bar{E}} \exp{[-i z/\lambdabar]}$, does not vary much along
$z$ on the scale of the reduced wavelength $\lambdabar$. With some
abuse of language we will indicate $\vec{\widetilde{E}}$ as "the
field". The field obeys the following paraxial wave equation in
the space-frequency domain: $ \mathcal{D}
\left[\vec{\widetilde{E}}(z,\vec{r},\omega)\right] = \vec{g}(z,
\vec{r},\omega)$. Here $\omega=2 \pi c/\lambda$, and the
differential operator $\mathcal{D}$ is defined by $\mathcal{D}
\equiv \left({\nabla_\bot}^2 + {2 i \omega /{c}} \cdot
{\partial_z}\right)$, ${\nabla_\bot}^2$ being the Laplacian
operator over transverse cartesian coordinates. The source-term
vector $\vec{g}(z, \vec{r})$ is specified by the trajectory of the
source electrons, and can be written in terms of the Fourier
transform of the transverse current density, $\vec{\bar{j}}
(z,\vec{r},\omega)$, and of the charge density,
$\bar{\rho}(z,\vec{r},\omega)$, as $\vec{g} = - {4 \pi}
\exp\left[-{i \omega z}/{c}\right] \left({i\omega}/{c^2} \cdot
\vec{\bar{j}} -\vec{\nabla}_\bot \bar{\rho}\right)$. Vector
$\vec{\bar{j}}$ and $\bar{\rho}$ are regarded as given data. In
this paper we will treat $\vec{\bar{j}}_\bot$ and $\bar{\rho}$ as
macroscopic quantities, without investigating individual electron
contributions.

%The part of the setup preceding the collimation section does not
%contribute to the field detected after the beam dump. In fact, the
%small amount of density modulation before the collimator is
%responsible for electromagnetic-source harmonics that are
%vertically shifted with respect to the electromagnetic-source
%harmonics after the collimator. It follows that the wave packets
%produced before and after the collimation section can be treated
%independently. In particular, we are interested in the field
%produced after collimation.

From the previous discussion it follows that, as concerns emission
of coherent optical radiation, our setup reduces to an upstream
bending magnet (corresponding to the last bend of the collimator)
followed by a straight section, an undulator (the main SASE 1
undulator), a second straight section and a downstream separation
bending magnet, which divides the electron beam from the XFEL
pulse (see Fig. \ref{realscheme}). We picture the upstream bending
magnet as a "switch-on" of both harmonics of the electromagnetic
sources and of the field. Similarly, the downstream bend enforces
a "switch-off" process.

When a modulated beam passes through a straight section limited by
upstream and downstream bending magnets, it produces edge
radiation (see among others, \cite{BOSF,CHUN} and references
therein). In our case, trajectory from the collimator section up
to the beam dump is more complicated than a single straight
section limited by bending magnets, but conceptually the mechanism
of radiation production is the same. Moreover, the influence of
bending magnet radiation to the field contribution can be shown to
be negligible. To see this, it is sufficient to compare the
radiation formation length of the field associated with bending
magnets with the radiation formation length of the shortest
straight section. Deflection introduced in the collimation section
corresponds to a bending radius of $400$ m. Deflection introduced
between SASE 1 and SASE 3 corresponds, instead, to a bending
radius of about $200$ m. For a wavelength $\lambda = 400$ nm and
bending radius $R= 400$ m we obtain the longest formation length
$(\lambdabar R^2)^{1/3} \simeq 0.2$ m. For the shortest straight
section of length $L = 200$ m, the edge-radiation formation length
would be $\min[L, 2 \gamma^2\lambdabar] = 2 \gamma^2\lambdabar
\simeq 150$ m, which is about $10^3$ times longer than the
formation length for the bends. It follows that field
contributions from bending magnets can be neglected with an
accuracy $10^{-3}$, and a sharp-edge approximation applies.

Understanding the operation of the optical radiator is made
simpler with the help of the theoretical study in reference
\cite{OURF}. In that reference we showed that radiation from an
ultra-relativistic filament beam along the trajectory specified
above can be interpreted as radiation from three virtual sources,
located at specific longitudinal positions. Specification of these
virtual sources amounts to specification of three field
distributions $\vec{\bar{E}}(\vec{r},z_s)$ at three distinct
locations $z_s$. In principle these locations are a matter of
taste, but there are particular choices of $z_s$ where
$\vec{\bar{E}}(\vec{r},z_s)$ exhibit plane wavefronts and are
similar to waists of laser beams. In the case under study these
privileged longitudinal positions are the center of the straight
sections and the center of the undulator. Let us indicate with
$L_1$ and $L_2$ the lengths of upstream and downstream straight
sections, with $L_w$ the length of the main XFEL undulator and set
$z=0$ in its center. In this case, the three sources are located
at $z_{s1}=-(L_w+L_1)/2$, $z_{s2}=0$ and $z_{s3} = (L_w+L_2)/2$.
Since the virtual sources exhibit plane wavefronts, they are
completely specified by real-valued amplitude distributions of the
field. In the case of a single electron, these were derived from
the far zone field distribution and were found to be \cite{OURF}:

\begin{eqnarray}
\vec{\widetilde{E}}_{1}\left(z_{s1},\vec{r}\right) &&= -
\exp\left[-\frac{i }{4}\left(\frac{L_w}{\gamma_z^2 \lambdabar
}+\frac{L_1}{\gamma^2 \lambdabar}\right)\right]\cr && \times
\frac{e L_1}{2\pi c \lambdabar^2 }  \int d\vec{\theta}
~\vec{\theta}~ \mathrm{sinc}\left[\frac{1}{4} \left(\frac{L_1
\theta^2}{\lambdabar}+\frac{L_1}{\gamma^2 \lambdabar}\right)
\right]\exp\left[\frac{i}{\lambdabar}\vec{r}\cdot \vec{\theta}
\right] ~,\label{vir1eaf}
\end{eqnarray}
\begin{eqnarray}
\vec{\widetilde{E}}_{2}(z_{s2},\vec{r}) &=& - \frac{e L_w}{2\pi c
\lambdabar^2 } \int d\vec{\theta} ~\vec{\theta}~
\mathrm{sinc}\left[\frac{1}{4} \left(\frac{L_w
\theta^2}{\lambdabar}+\frac{L_w}{\gamma_z^2 \lambdabar}\right)
\right]\exp\left[\frac{i}{\lambdabar}\vec{r}\cdot \vec{\theta}
\right] ~\cr && \label{vir1ebf}
\end{eqnarray}
and

\begin{eqnarray}
\vec{\widetilde{E}}_{3}\left(z_{s3},\vec{r}\right) &&= -
\exp\left[\frac{i }{4}\left(\frac{L_w}{\lambdabar
\gamma_z^2}+\frac{L_2}{\gamma^2 \lambdabar }\right)\right]\cr &&
\times \frac{e L_2}{2\pi c \lambdabar^2 }  \int d\vec{\theta}
~\vec{\theta}~ \mathrm{sinc}\left[\frac{1}{4} \left(\frac{L_2
\theta^2}{\lambdabar}+\frac{L_2}{\gamma^2 \lambdabar}\right)
\right]\exp\left[\frac{i}{\lambdabar}\vec{r}\cdot \vec{\theta}
\right] ~.\label{vir1edf}
\end{eqnarray}
Note that the field is radially polarized. Aside for a different
phase, Eq. (\ref{vir1eaf}), Eq. (\ref{vir1ebf}) and Eq.
(\ref{vir1edf}) have a similar mathematical structure. However,
for SASE 1, $K=3.3$, so that $\gamma_z^2 = \gamma^2/(1+K^2/2)$ is
about an order of magnitude smaller than $\gamma^2$. It follows
that the second term in the $\mathrm{sinc}(\cdot)$ function of Eq.
(\ref{vir1eaf}) or Eq. (\ref{vir1edf}) is much smaller than the
analogous term in Eq. (\ref{vir1ebf}). In particular, for our
study case we have ${L_1}/({4 \lambdabar \gamma^2}) = 0.97$ and
${L_2}/({4\lambdabar \gamma^2}) = 0.67$ of order unity, but
${L_w}/({4\lambdabar \gamma_z^2}) = 4.3$. As a result, the
$\mathrm{sinc}(\cdot)$ function in Eq. (\ref{vir1edf}) is strongly
suppressed, one may neglect the virtual source located in the
center of the undulator and let
$\vec{\widetilde{E}}_{2}(z_{s2},\vec{r})=0$, at least in first
approximation. It should be remarked that the virtual source
$\vec{\widetilde{E}}_{2}$ is responsible for a field contribution
known as transition undulator radiation. Typical expressions for
TUR emission \cite{KIM1,KINC,CAST} consist of relations for the
energy distribution of radiation in the far zone that account for
the presence of the virtual source at $z_{s2}$ alone, without
considering further contributions due to other elements in the
beamline, e.g. the straight sections in our case. As it was shown
in \cite{BOS4} and \cite{OURF} these expressions have no physical
meaning. In our study case, as we have just seen, the contribution
from transition undulator radiation can be even neglected.

We are thus left with the contributions from two virtual sources
located at $z_{s1}$ and $z_{s3}$, accounting for edge radiation
emission from the straight lines before and after the undulator.
However, the contribution from the upstream straight section will
be strongly suppressed by a photon stop inside the main undulator,
whose main function is that of absorbing spontaneous radiation
background. Moreover, an extra photon stop might be installed at
the exit of the main undulator, absorbing all but the SASE pulse.
As a result, one has to deal with the simple situation where the
optical radiator is composed by a single straight section limited
by bending magnets downstream of the main undulator. Since
$\vec{\bar{E}}(\vec{r},z_{s3})$ is known, free-space propagation
from the virtual source through the near zone and up to the
far-zone can be performed with the help of the Fresnel propagation
as it is done for usual laser beams \cite{ARTF}. The only
difference is in the peculiar shape of
$\vec{\bar{E}}(\vec{r},z_{s3})$, which reflects the particular
trajectory followed by the filament beam. It is important to
realize that the first optical element of the optical beamline
will be placed at about $300$ m from the end of the straight
section. Such distance is comparable with the straight section
length. Thus, one needs to know how edge radiation propagates in
the near zone in order to characterize the pulse at the optical
element position. In this regard, it should be stressed that
Fresnel propagation allows one to calculate the electric field not
only in the far zone, but in the near zone too, solving the full
problem of free-space propagation.

In principle, one may directly propagate Eq. (\ref{vir1edf}).
However, the integral in $d\vec{\theta}$ cannot be solved
analytically. Numerical evaluations are simplified with the help
of reference \cite{OURF}, where we showed that the virtual source
in Eq. (\ref{vir1edf}) is equivalent to two virtual sources
located at the edges of the straight section. These two sources
still present a plane wavefront, and they can be described
analytically in a simple way in terms of the modified Bessel
function of the first order $K_1(\cdot)$. It is convenient to
adopt this picture for computational purposes. Shifting, for
simplicity, the origin of the longitudinal axis in the center of
the downstream straight section and letting $L \equiv L_2$ we can
write the two virtual sources as \cite{OURF}:

\begin{figure}
\begin{center}
\includegraphics*[width=140mm]{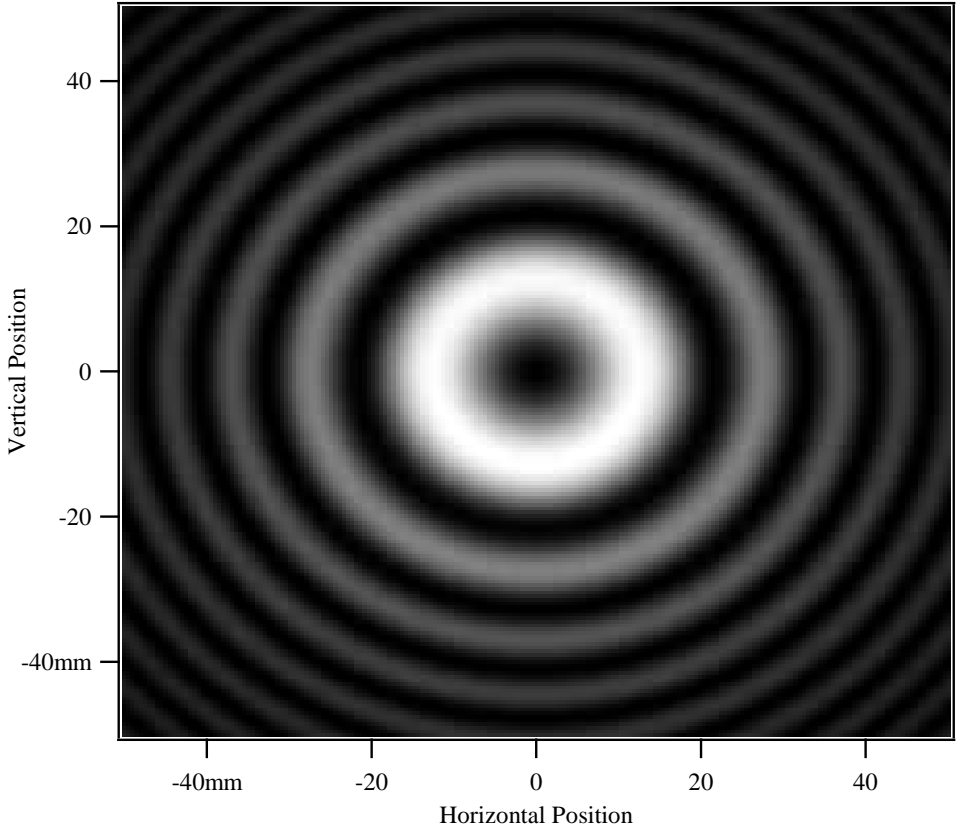}% Here is how to import EPS art
\caption{\label{SRW2D} Contour plot of the spatial distribution of
radiation simulated with the help of the SRW code at the position
of the first optical element, $z=400$ m. }
\end{center}
\end{figure}
\begin{figure}
\begin{center}
\includegraphics*[width=140mm]{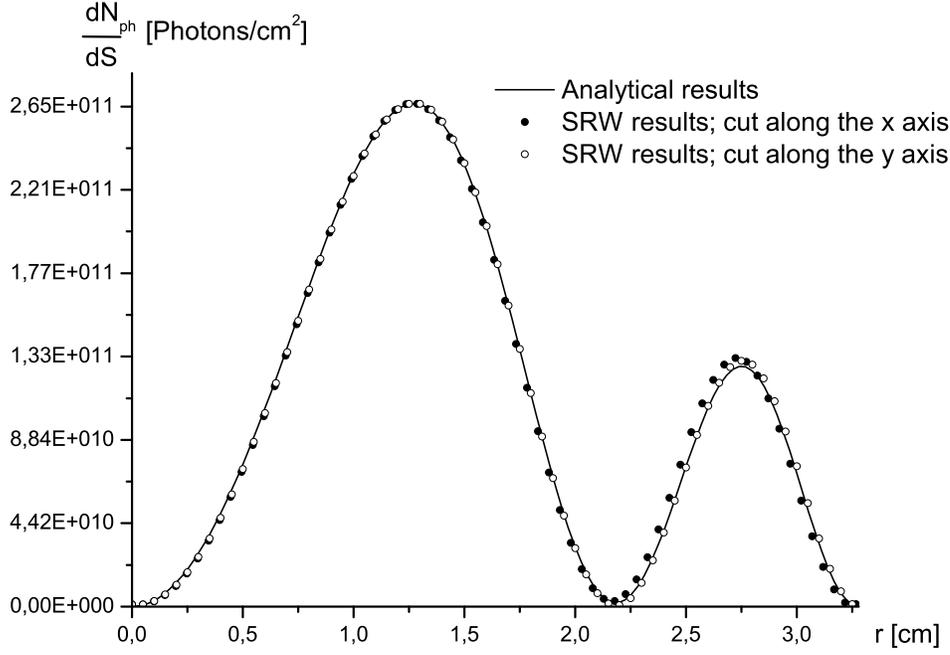}% Here is how to import EPS art
\caption{\label{profile} Photon density distribution as a function
of the radial position at the first optical element, $z=400$ m.
The solid line has been calculated with the help of Eq.
(\ref{fiprF}). Black and white circles are obtained by cutting the
contour plot in Fig. \ref{SRW2D} respectively along the horizontal
and vertical axis.}
\end{center}
\end{figure}

\begin{eqnarray}
\vec{\widetilde{E}}\left(\pm \frac{L}{2},\vec{r}\right) = \mp
\frac{2 i e}{c \gamma \lambdabar} \exp\left[\pm \frac{i L }{4
\gamma^2  \lambdabar}\right] \frac{\vec{r}}{r}
K_1\left(\frac{r}{\gamma \lambdabar} \right)~.\label{virpm05}
\end{eqnarray}
Fresnel propagation can now be performed, and the radiation energy
density can be calculated as \cite{OURF}

\begin{eqnarray}
\frac{dW}{d\omega dS} &=& \frac{e^2}{4\pi^2 \lambdabar L c}
F\left(z,\vec{r}\right)~\label{fipr}
\end{eqnarray}
with

\begin{eqnarray}
F\left(z,\vec{r}\right) &=& \left| \left\{-\frac{\vec{r}}{r}
\left(\frac{2 i  }{{z}-L/2}\right){\frac{L}{\gamma\lambdabar
\sqrt{\lambdabar L}}} \exp\left[\frac{i r^2 }{2
\lambdabar\left({z}-L/2\right)}\right] \exp\left[\frac{i}{4}
\frac{L}{\gamma^2\lambdabar}\right] \right.\right.\cr &&
\left.\left. \times \int_0^{\infty} d{r}' {r}'
K_1\left(\frac{r'}{\gamma\lambdabar}\right) J_1
\left(\frac{r{r}'}{\lambdabar({z}-L/2)}\right)
\exp\left[\frac{i{r}'^2 }{2\lambdabar\left({z}-L/2\right)}\right]
\right\} \right.\cr &&\left. + \left\{\frac{\vec{r}}{r}
\left(\frac{2 i }{{z}+L/2}\right){\frac{L}{\gamma\lambdabar
\sqrt{\lambdabar L}}} \exp\left[\frac{i r^2 }{2
\lambdabar\left({z}+L/2\right)}\right] \exp\left[-\frac{i}{4}
\frac{L}{\gamma^2\lambdabar}\right] \right.\right.\cr &&
\left.\left. \times \int_0^{\infty} d{r}' {r}'
K_1\left(\frac{r'}{\gamma\lambdabar}\right) J_1
\left(\frac{r{r}'}{\lambdabar({z}+L/2)}\right)
\exp\left[\frac{i{r}'^2 }{2\lambdabar\left({z}+L/2\right)}\right]
\right\}  \right|^2~.\cr &&\label{fiprF}
\end{eqnarray}
Since we are interested in the radiation energy from a filament
beam with a given longitudinal profile, we should multiply the
single-electron result by the squared-modulus of the Fourier
transform of the temporal profile of the bunch. Near the
modulation frequency $c/\lambdabar$, a bunch with modulated
Gaussian temporal profile and rms duration $\sigma_T$ gives

\begin{eqnarray}
\bar{f}(\omega)=\frac{N
a_f}{2}\left\{\exp\left[-\frac{\sigma_T^2}{2}
\left(\omega-\frac{c}{\lambdabar}\right)^2\right]
+\exp\left[-\frac{\sigma_T^2}{2}
\left(\omega+\frac{c}{\lambdabar}\right)^2\right]\right\}~,
\label{fbarrrr}
\end{eqnarray}
where $\bar{f}(\omega)$ is the Fourier transform of the temporal
profile of the bunch and $N$ is the number of electrons in the
bunch.

In order to calculate the spatial density distribution of the
number of photons per pulse we should integrate Eq. (\ref{fipr})
in $d \omega$. Since we are interested in coherent emission around
the modulation wavelength, we can consider the wavelength in Eq.
(\ref{fipr}) fixed. This amounts to a multiplication of Eq.
(\ref{fipr}) by

\begin{eqnarray}
\int_0^\infty d\omega
\left|\bar{f}(\omega)\right|^2=\frac{\sqrt{\pi} N^2 a_f^2}{4
\sigma_T}~, \label{squarefbar}
\end{eqnarray}
leading to

\begin{eqnarray}
\frac{d N_{ph}}{dS} = \frac{\sqrt{\pi}}{16 \pi^2} \frac{N^2 \alpha
a_f^2}{c L \sigma_T} F\left(z,\vec{r}\right)~, \label{endenee}
\end{eqnarray}
where $\alpha \equiv e^2/(\hbar c) = 1/137$ is the fine structure
constant.

\begin{figure}
\begin{center}
\includegraphics*[width=140mm]{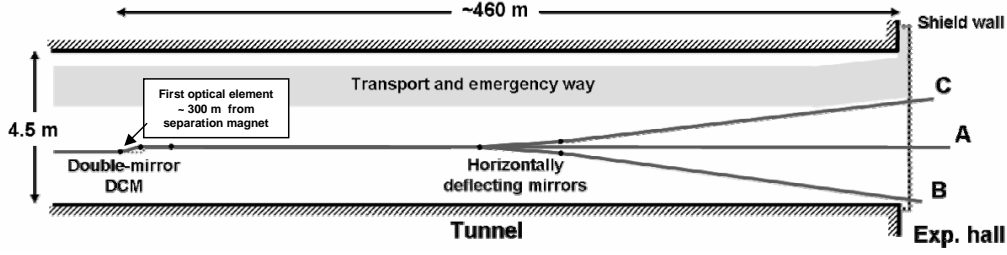}% Here is how to import EPS art
\caption{\label{tunnel} Schematic representation of the optical
transport system in the European XFEL.}
\end{center}
\end{figure}

We considered (see \cite{XFEL}) the case for $|a_f|=0.1$, $L=200
$m, $\sigma_T \simeq 80$ fs and $N \simeq 6 \cdot 10^9$ (i.e.
about $1$ nC). Since the first element of the optical beamline is
foreseen to be placed at about $300$ m from the separating magnet
(see Fig. \ref{realscheme} and Fig. \ref{tunnel}), and since we
measure $z$ from the center of the straight section, we set our
observation plane at $z = 400$ m. Then we use Eq. (\ref{endenee})
to calculate the photon density distribution. We cross-checked our
analytical results with the code SRW \cite{CHUB}. A contour plot
for the photon density distribution as calculated from SRW is
given in Fig. \ref{SRW2D}. Horizontal and vertical cuts along the
contour plot are compared with Eq. (\ref{endenee}) in Fig.
\ref{profile}. The total number of photons between the first two
minima of the distribution function (at $r=0$ cm and $r \simeq
2.2$ cm respectively) is obtained integrating in $dS$. For
parameters selected above, $N_{ph} \simeq 2 \cdot 10^{12}$
photons.

The optical pulse must be transported to the experimental area
preserving relative synchronization with the XFEL pulse. The most
convenient way to accomplish this task is to use the same physical
support for both x-ray and edge-radiation optics, e.g. assembling
mirrors for the transport of edge radiation directly on x-ray
optical elements. In this way, mechanical vibrations of the
transport system will not affect relative synchronization of the
two pulses. A scheme of the XFEL optical system is sketched in
Fig. \ref{tunnel}. As mentioned before, the first optical elements
will be located about $300$ m downstream of the separation magnet.
Once edge-radiation is transported  to the experimental area,
single-shot cross-correlation with the pump-laser pulse can be
performed, yielding the time delay between the pump-laser and the
XFEL pulse. As discussed before, in reference \cite{CROS} a
measurement of the time-offset signal between two ultrashort laser
pulses was experimentally performed based on sum-frequency
generation in a non-linear crystal. The intrinsic accuracy of the
method is in the femtosecond range. Actual measurements were
successfully performed even when one of the two laser pulses is
very weak, up to $10^{6}$ photons per pulse. This sensitivity
allows our method to be employed even if a large number of optical
photons are lost, from the total of $10^{12}$ photons per pulse,
during the transport process along the optical system.

\section{\label{sec:conc} Conclusions}

Our analysis demonstrates the feasibility of pump-probe
experiments at XFELs with femtosecond temporal resolution, based
on timing of XFEL pulses to optical pulses from an external
pump-laser. The proposed scheme does not require absolute
synchronization of pump and probe pulses. Synchronization in the
sub-picosecond range, which has been experimentally demonstrated,
is sufficient for its operation.

The present study includes an analysis of physical principles of
operation, which are clear and transparent, and of fundamental
physical effects of importance for the operation of the proposed
time-arrival monitor.

Technical realization will be rather simple and cost-effective
since it is essentially based on technical components
(optical-replica synthesizer) being already included in the design
of the European XFEL. Baseline parameters of the European XFEL
were used in our analysis. Thus, the proposed scheme can be
implemented at the first stage of XFEL facility operation.

\section*{\label{sec:graz} Acknowledgements}

We thank  Massimo Altarelli (DESY) and Jochen Schneider (DESY) for
their interest in this work, Vitali Kocharian (DESY) for his help
with SRW simulations.


\begin{thebibliography}{99}

\bibitem{Zewail}
A.H. Zewail, J. Phys. Chem. A, 104 (2000) 5660

\bibitem{XFEL} M. Altarelli et al. (Eds.), XFEL: The European X-Ray Free-Electron
Laser. Technical Design Report, DESY 2006-097, DESY, Hamburg
(2006) (See also http://xfel.desy.de).

\bibitem{SLAC} J. Arthur et al. Linac Coherent Light Source
(LCLS). Conceptual Design Report, SLAC-R593, Stanford (2002) (See
also http://www-ssrl.slac.stanford.edu/lcls/cdr).

\bibitem{SCSS}
Tanaka, T. \& Shintake, T. (Eds.): SCSS X-FEL Conceptual Design
Report. Riken Harima Institute, Hyogo, Japan, May 2005 (see also
http://www-xfel.spring8.or.jp).

\bibitem{GS1}
A. Winter et al., Proc. 27th Int. FEL Conference (Stanford, 2005),
p.676.

\bibitem{GS2}
J. Kim et al., Proc. 28th Int. FEL Conference (Berlin, 2006),
p.287.

\bibitem{GS3}
J. Kim et al., Proc. EPAC 2006 (Edinburgh, 2006), p.2744.

\bibitem{TC1}
B. Faatz et al., Nucl. Instrum. and Meth. in Phys. Res. A 475
(2001), 363


\bibitem{TC2}
J. Feldhaus et al., Nucl. Instrum. and Meth. in Phys. Res. A 528
(2004), 453

\bibitem{fir-pp}
M. Gensch et al., New infrared undulator beamline at FLASH,
Infrared Physics and Technology, in press.

\bibitem{BILL} A. A. Zholents and W. M. Fawley, Phys. Rev. Lett.
92, 224801 (2004)

\bibitem{fel07-TUPPH013}
E.L.~Saldin, E.A.~Schneidmiller and M.V.~Yurkov, Proc. 29th Int.
FEL Conference (Novosibirsk, 2007), TUPPH013

\bibitem{ORS1} E. Saldin, E. Schneidmiller and M. Yurkov, Nucl.
Instrum. and Meth. in Phys. Res. A 539 (2005), 499

\bibitem{CROS} V. Tenishev, V. Avdeichikov, A. Persson and J.
Larsson, Meas. Sci. Technol. 15 (2004), 1762

\bibitem{Czonka}
P. Czonka, Part. Accel. 80 (1978) 225

\bibitem{LSCF} E. Saldin, E. Schneidmiller and M. Yurkov, Nucl.
Instrum. and Meth. in Phys. Res. A 528 (2004), 335

\bibitem{ROSE} J. Rosenzweig et al. in Proc. of Advanced Accelerator
Workshop, Lake Tahoe (1996)

\bibitem{LWF1} G. Geloni, E. Saldin, E. Schneidmiller and M. Yurkov, Nucl.
Instrum. and Meth. in Phys. Res. A 578, 1 (2007), 34

\bibitem{SCTH} G. Geloni, E. Saldin, E. Schneidmiller and M. Yurkov, Nucl.
Instrum. and Meth. in Phys. Res. A 554 (2005), 20

\bibitem{OURX} G. Geloni, E. Saldin, E. Schneidmiller and M. Yurkov, Nucl.
Instrum. and Meth. in Phys. Res. A 583 (2007), 228

\bibitem{FAST} E.L. Saldin, E.A. Schneidmiller, M.V. Yurkov, Nucl.
Instr. and Meth. A 429 (1999), 233

\bibitem{ARTF} G. Geloni, E. Saldin, E. Schneidmiller and M.
Yurkov, Optics Communications 276, 1 (2007), 167

\bibitem{BOSF} R.A. Bosch et al. Rev. Sci. Instrum. 67, 3346 (1995)

\bibitem{CHUN} R.A. Bosch and O.V. Chubar, "Long wavelength edge
radiation in an electron storage ring", Proc. SRI'97, tenth US
National Conference, AIP Conf. Proc. 417, edited by E. Fontes
(AIP, Woodbury, NY), pp. 35-41 (1997)

\bibitem{OURF} G. Geloni, E. Saldin, E. Schneidmiller and M.
Yurkov, "Fourier Optics Treatment of Classical Relativistic
Electrodynamics", DESY 06-127 (2006) (see also
http://arxiv.org/abs/physics/0608145)

\bibitem{KIM1} K.-J. Kim, Phys. Rev. Lett. 76, 8 (1996)

\bibitem{KINC} B.M. Kincaid, Il Nuovo Cimento Soc. Ital. Fis, 20D,
495 (1998) and LBL-38245 (1996)

\bibitem{CAST} M. Castellano, Nucl. Instr and Meth. in Phys. Res.
A 391 (1997) 375

\bibitem{BOS4} R.A. Bosch, Il Nuovo Cimento, 20, 4 p. 483 (1998)

\bibitem{CHUB} O.Chubar and P.Elleaume, in Proc. of the 6th European Particle Accelerator Conference EPAC-98, 1177-1179 (1998)


\end{thebibliography}
\end{document}